\documentclass[prd,nofootinbib,12pt]{revtex4-1}
\DeclareRobustCommand{\baselinestretch{1.3}}

\usepackage{amssymb,amsthm}

\def\bk{{\bf k}}
\def\bn{{\bf n}}
\def\bx{{\bf x}}

\def\im{{\rm i}}

\def\d3#1{\lower-1pt\hbox{$\stackrel{\ldots}{#1}$}}

\begin{document}

\title{{Erratum: Statistical anisotropy in the inflationary universe} \\ {[Phys. Rev. D 80, 023521
(2009)]}}

\author{Yuri Shtanov}\email{shtanov@bitp.kiev.ua}
\author{Hanna Pyatkovska}

\begin{abstract}
An error was made when calculating the amplitude of the perturbed scalar mode after
Hubble-radius crossing.  When corrected, it eliminates the leading term in the
statistical anisotropy of the power spectrum, reducing it from the previously claimed
upper estimate of $\lesssim 10^{-2}$ to the level of at most $10^{-6}$, which is beyond
the possibility of detection.  The general conclusion is that a single-field inflationary
scenario cannot produce statistical anisotropy of appreciable magnitude.
\end{abstract}


\maketitle

An error was made when calculating the amplitude of the perturbed mode after
Hubble-radius crossing, i.e., in passing from (85) to (90) and (91).  An important
circumstance was overlooked, namely, that the effective wave number and frequency of the
mode (85) are slightly different from the corresponding quantities ($\bk$ and $k$) of the
unperturbed mode, being equal to $\bk - \delta \bk$ and $k - \delta k$, respectively,
where $\delta \bk$ is the same as in the calculation of Appendix~A and is given by (A4)
for the contribution from tensor inhomogeneity. This should be taken into account when
using the condition of Hubble-radius crossing for this mode. The corrected equations (90)
and (91) will read
$$
A_\bk ( \bx ) = A_{k - \delta k} \frac{(k - \delta k)^{3/2}}{k^{3/2}}
e^{- \im \widetilde S_\bk \left(\tau_k, \bx \right) - \im \widetilde T_\bk \left(\tau_k,
\bx \right)} \, , \eqno(90)
$$
and
$$
\Phi_\bk \left( \tau_f, \bx \right) \approx - A_{k - \delta k} \frac{(k - \delta k)^{3/2}}{k^{3/2}}
e^{- \im \widetilde S_\bk \left(\tau_k, \bx \right) - \im \widetilde T_\bk \left(\tau_k,
\bx \right)} \, , \eqno(91)
$$
respectively.

Consequently, it is necessary to make an additional correction for the amplitude
$|A_k|^2$ in (A8), described by the substitution
$$
|A_k|^2 \to |A_{k - \delta k}|^2 \frac{(k - \delta k)^3}{k^3} = | A_k|^2 \left[ 1 -
\frac{d \ln \left( k^3 |A_k|^2 \right)}{d \ln k} \frac{\delta k}{k} \right] \, ,
$$
which will cancel the first term in (A9). Equations (A9) and (A11) will then read
$$
\mu_\bk^T = Q + \Bigl[ \bn \cdot \nabla Q - \left( \bn \cdot \nabla \right)^2 T_0 \Bigr]
\frac{d \tau_k}{d \ln k} \, . \eqno{({\rm A}9)}
$$
$$
\nu_\bk^T = \frac12 \left(\mu_\bk^T + \mu_{-\bk}^T \right) =  Q - \frac12 \left(\bn \cdot
\nabla \right)^2 \Bigl[ (T_0)_\bk + (T_0)_{-\bk} \Bigr] \frac{d \tau_k}{d \ln k} \, . \eqno{({\rm A}11)}
$$

A similar cancellation will take place in the contribution $\nu_\bk^S$ in (93).  The
leading terms written in (94) and (95) are then absent, and the numerical estimates of
Secs.~V--VI and in the Summary Sec.~VIII are then of no relevance. The corrected
expressions read
$$
\nu_\bk^S =  - \frac12 \left(\bn \cdot \nabla \right)^2 \Bigl[ (S_0)_\bk + (S_0)_{-\bk}
\Bigr] \frac{d \tau_k}{d \ln k}  \, , \eqno{(95)}
$$
$$
\nu_\bk^T =  Q - \frac12 \left(\bn \cdot \nabla \right)^2 \Bigl[ (T_0)_\bk + (T_0)_{-\bk}
\Bigr] \frac{d \tau_k}{d \ln k} \, .  \eqno{(96)}
$$

The estimates of the terms in (95) and (96) above, made in Appendix~A [see Eq.~(A12) and
the paragraph that follows this equation], imply that the generated anisotropy $\nu_\bk =
\nu_\bk^S + \nu_\bk^T$ of the primordial power spectrum in (122), instead of the claimed
upper value $\nu_\bk \lesssim 10^{-2}$, is at most of the order $10^{-6}$, which is too
small to be detectable.

The general conclusion is that a single-field inflationary scenario cannot produce
statistical anisotropy of appreciable magnitude.

\end{document}